%% file: main.tex
\documentclass[sigconf,nonacm]{acmart}

\usepackage{balance}
\usepackage{bm}
\usepackage{threeparttable}
\usepackage{booktabs} 
\usepackage{multirow}
\usepackage{subfigure}
\usepackage{enumitem}
\usepackage{balance}
\usepackage{diagbox}
\usepackage{enumitem}
\definecolor{lightgreen}{rgb}{0.4, 0.8, 0.4}
\setlist[itemize]{leftmargin=*}

\AtBeginDocument{%
  }

\setcopyright{acmlicensed}
\copyrightyear{2018}
\acmYear{2018}
\acmDOI{XXXXXXX.XXXXXXX}
\acmConference[Conference acronym 'XX]{Make sure to enter the correct
  conference title from your rights confirmation emai}{June 03--05,
  2018}{Woodstock, NY}
\acmISBN{978-1-4503-XXXX-X/18/06}

\begin{document}

\title{Heterogeneous Multi-treatment Uplift Modeling for Trade-off Optimization in Short-Video Recommendation}

\author{Chenhao Zhai}
\authornote{Both authors contributed equally to this research.}
\authornote{Work done when he was a research intern at Kuaishou Technology.}
\affiliation{
  \institution{Kuaishou Technology}
  \city{Beijing}
  \country{China}
}
\email{dch23@mails.tsinghua.edu.cn}

\author{Chang Meng}
\authornotemark[1]
\affiliation{
  \institution{Kuaishou Technology}
  \city{Beijing}
  \country{China}
}
\email{mengchang@kuaishou.com}

\author{Xueliang Wang}
\affiliation{
  \institution{Kuaishou Technology}
  \city{Beijing}
  \country{China}
}
\email{wangxueliang03@kuaishou.com}

\author{Shuchang Liu}
\affiliation{
  \institution{Kuaishou Technology}
  \city{Beijing}
  \country{China}
}
\email{liushuchang@kuaishou.com}

\author{Xiaolong Hu}
\affiliation{
  \institution{Kuaishou Technology}
  \city{Beijing}
  \country{China}
}
\email{huxiaolong@kuaishou.com}

\author{Shisong Tang}
\affiliation{
  \institution{Kuaishou Technology}
  \city{Beijing}
  \country{China}
}
\email{tangshisong@kuaishou.com}

\author{Xiaoqiang Feng}
\authornote{The corresponding author.}
\affiliation{
  \institution{Kuaishou Technology}
  \city{Beijing}
  \country{China}
}
\email{fengxiaoqiang@kuaishou.com}


\author{Xiu Li}
\authornotemark[3]
\affiliation{
  \institution{Shenzhen International Graduate School, Tsinghua University}
  \city{Shenzhen}
  \country{China}
}
\email{li.xiu@sz.tsinghua.edu.cn}


\begin{abstract}
The rapid proliferation of short videos on social media platforms presents unique challenges and opportunities for recommendation systems. Users exhibit diverse preferences, and the responses resulting from different strategies often conflict with one another, potentially exhibiting inverse correlations between metrics such as watch time and video view counts. Existing uplift models face limitations in handling the heterogeneous multi-treatment scenarios of short-video recommendations, often failing to effectively capture both the synergistic and individual causal effects of different strategies. Furthermore, traditional fixed-weight approaches for balancing these responses lack personalization and can result in biased decision-making. To address these issues, we propose a novel Heterogeneous Multi-treatment Uplift Modeling (HMUM) framework for trade-off optimization in short-video recommendations. HMUM comprises an Offline Hybrid Uplift Modeling (HUM) module, which captures the synergistic and individual effects of multiple strategies, and an Online Dynamic Decision-Making (DDM) module, which estimates the value weights of different user responses in real-time for personalized decision-making. Evaluated on two public datasets, an industrial dataset, and through online A/B experiments on the Kuaishou platform, our model demonstrated superior offline performance and significant improvements in key metrics. It is now fully deployed on the platform, benefiting hundreds of millions of users. 
\end{abstract}


\maketitle

\input{sections/introduction}
\input{sections/related_work}
\input{sections/preliminary}
\input{sections/method}
\input{sections/experiments}

\input{sections/application}
\input{sections/conclusion}


\bibliographystyle{ACM-Reference-Format}
\bibliography{main}

\appendix

\input{sections/appendix.tex}









\end{document}

%% file: sections/introduction.tex
\section{INTRODUCTION}

Video recommendation platforms such as YouTube, TikTok, and Kuaishou have become popular in people's social life for their ability to fulfill users' diverse and dynamic information needs. 
While the mainstream recommendation methods \cite{cuicalibrating,wu2025aligning} focus on designing better ranking models \cite{liu2009learning, ricci2021recommender,li2024contextual,li2025contrastive,tang2025retrieval} (i.e., learning to sort and select the candidate videos that better aligns with the user's preferences), the industrial solution often engage a series of complementary strategies (or modules) that controls the recommendation statistics to circumvent model biases.




In our industrial system, we have implemented several straightforward solutions. 
The general framework leverages statistical analyses of user behavior data as individual correction factors to adjust the parameters of specific strategies. 
While these naive approaches have demonstrated empirical effectiveness, they often either fail to capture the dynamic causal relationships between the decision-making process and user behavior statistics nor neglect the synergy between different strategies. 
To achieve improved generalization performance, we require a robust causal modeling framework that can effectively explore and adapt to user interests.

To tackle the task at hand, we propose reformulating the problem using uplift modeling. 
This personalized causal effect modeling technique is commonly employed in online marketing scenarios, such as distributing coupons and discounts. 
The primary function of uplift modeling is to measure the incremental impact (i.e., uplift) of treatments or interventions on different populations \cite{efin,ganite,cdum,li2025klan,sun2024end}, enabling data-driven decision-making. 
Representative methods include meta-learning models such as T-learner \cite{metalearners} and S-learner \cite{metalearners}, as well as tree-based models \cite{wang2015robust,radcliffe2011real,athey2016recursive, nandy2023generalized} and Deep Uplift Modeling (DUM) \cite{he2024rankability,huang2024entire,bnn,euen,liu2024benchmarking,xu2022learning,curth2021nonparametric} . 
Drawing inspiration from the success of uplift modeling in online marketing, we reformulate the trade-off optimization problem in our video recommendation pipeline as a multi-treatment dynamic uplift modeling task.

Nevertheless, applying the uplift modeling in the video recommendation scenario presents additional challenges:
\begin{itemize}
    \item \textbf{Effective uplift modeling for heterogeneous multiple treatment in video recommendation scenario.} 
    Unlike in online marketing scenarios, in the short video context, different strategies can be viewed as multiple treatments. 
    These strategies are heterogeneous, each exerting diverse and often mutually exclusive effects on user responses.    
    Existing uplift modeling approaches for the same response can be categorized into two types. 
    The first type \cite{euen,efin,sun2024m} employs a single model to jointly model multiple treatments, effectively identifying dependencies and synergies between them. 
    However, it may overlook the individual causal effects of single specific treatment. 
    More critically, these methods are designed for \emph{homogeneous} multi-treatment scenarios (e.g., treatments with varying intensities within the same scope). 
    However, since the effects of treatments on user responses are often \emph{heterogeneous} and mutually exclusive, these joint-learning approaches can lead to significant gradient conflicts, potentially compromising the model's effectiveness. 
    The second type \cite{descn,dragonnet} focuses on modeling the causal effect of a specific treatment. 
    While this avoids the complexity of interactions between heterogeneous treatments, it may fail to capture potential collaborative effects that could exist between them. 
    Considering the unique characteristics of short-video recommendation scenarios, characterized by heterogeneous multi-treatment, there is an urgent need for a solution that can effectively capture both the synergistic effects between treatments and the individual causal effects of each treatment.
    \item \textbf{Decision-making for trade-offs among multiple responses.} 
    In the short-video recommendation scenario, there are significant distribution differences among various responses, such as watch time (which focuses on user consumption-side feedback, i.e., consumption-side score) and video view counts (which focuses on user commercialization-side feedback, i.e., commercialization-side score). 
    Balancing the importance of these different metrics involves trade-offs. 
    Traditionally, fixed weight parameters have been used to balance these responses, but this approach lacks personalization and may lead to biased decision-making, as users have diverse habits and preferences when watching short videos. 
    Therefore, it is crucial to develop personalized adjustment methods that can effectively model the joint benefits across multiple responses, taking into account individual user behaviors and preferences.
\end{itemize}

To address these two challenges, we carefully design a Heterogeneous Multi-treatment Uplift Modeling for Trade-off Optimization in Short-Video Recommendation (HMUM), which contains two parts, Offline Hybrid Uplift Modeling (HUM) module and Online Dynamic Decision-Making (DDM) module. 

To address the first challenge, HUM employs a hybrid paradigm designed to capture both the synergistic effects between treatments and the individual causal effects of each treatment. 
Specifically, it begins with a feature selection module that extracts relevant features to serve as inputs for each branch of the individual causal effects estimation module. 
HUM then estimates individual responses for each treatment in its respective branch, while using a multi-task learning structure to uncover synergistic effects between treatments. 
Crucially, to mitigate the bias caused by the varying distribution of responses estimated across different branches and to accurately model these responses, HUM employs KL divergence to constrain the representations induced by non-treatment across branches. 

To handle the second challenge, DDM primarily constructs an online value evaluation module to assess the value proportions of different types of responses generated by different users. 
To be specific, DDM constructs a multi-task learning structure which leverages the rich and varied features of users (e.g., gender, ages, commercialization value.) 
to predict the value weights corresponding to multiple user responses in real-time. Further, these scores are used as weights to aggregate and weight various types of user responses. 
The aggregated score is then used for the final decision-making, determining whether to apply or enhance a particular treatment for the user.

In summary, our works have the following contributions:
\begin{itemize}
    \item We highlight the challenges of uplift modeling in video recommendation scenarios, specifically regarding the uplift modeling for multiple treatments and online decision making between multiple responses. 
    We also provide a complete solution that can be applied to online platforms based in video recommendation scenarios.
    \item We introduce an innovative approach called Heterogeneous Multi-treatment Uplift Modeling for Trade-off Optimization in Short-Video Recommendation (HMUM). 
    This method simultaneously learns the synergistic and individual effects of different treatments through an Offline Hybrid Uplift Modeling (HUM) module. 
    Additionally, it employs an Online Dynamic Decision-Making (DDM) module to dynamically estimate the value weights of different responses for each user, delivering a balanced comprehensive score for the final online decision-making process.
    \item We evaluated our model on two public datasets, an industrial dataset, and through online A/B experiments on the Kuaishou platform.
    Our model not only exhibited superior offline performance but also achieved significant improvements in consumption metrics on A/B test. 
    Ultimately, our model was deployed on the online platform, where it now serves hundreds of millions of users.
    \item We plan to release the source code and the collected industrial dataset. 
    This will support researchers in further exploring the uplift modeling problem in video recommendation scenarios and contribute to the advancement of the community as a whole.
\end{itemize}

%% file: sections/related_work.tex
\section{RELATED WORK}
Uplift models aim to identify the target user groups for each specific treatment by accurately estimating the individual treatment effects (ITE). Traditional Uplift modeling methods primarily consist of meta-learner based methods and tree-based methods.

Meta-learner based methods \cite{yao2021survey,metalearners} are a classic uplift modeling framework. S-learner \cite{metalearners} and T-learner \cite{metalearners} are two commonly used meta-learning paradigms. The S-learner employs a single model to merge data from the treatment and control groups for modeling, and incorporates the treatment variable as a model input feature. The T-learner constructs separate models for the control and treatment groups, estimating ITE through the difference in model predictions. These paradigms are highly sensitive to data. That is, if the contribution of the treatment feature in the model is low or there is a significant imbalance in the amount of data between the treatment group and the control group, it can significantly affect the model performance and thus affect the estimation of ITE. To address this problem, the X-learner method \cite{metalearners} improves upon the T-learner by introducing the concept of propensity scores and two additional ITE estimators, combining propensity scores to weight the estimation of ITE. It makes full use of all data for prediction and has a better handling effect on the imbalance issue between the treatment and control groups. Tree-based methods \cite{wang2015robust,radcliffe2011real,athey2016recursive, nandy2023generalized} utilize specific tree or forest structures, setting direct computational methods for incremental Uplift modeling. They measure the difference before and after splitting to strategically split feature points, gradually dividing the entire space, and ultimately identifying a small number of sensitive sub-group users from the entire population.

With the development of deep learning, researchers have found that employing neural networks can introduce more complex and flexible architectures to model the response process to treatment, enabling the learning of more accurate ITE estimates. Therefore, such methods have become a popular framework for uplift modeling \cite{wei2024multi,cevae,bica2020estimating,sun2024towards,zhou2023direct}. Existing neural network-based Uplift modeling methods can be categorized into two categories: joint modeling of treatment and modeling of specific treatment causal effects. The former estimates the non-treatment score through features and incorporates the treatment feature into the treatment modeling based on this score. For example, EUEN \cite{euen} uses different sub-neural networks to estimate the conversion probability of the control group and the treatment causal effect. EFIN \cite{efin} captures the feature subset related to the control group through explicit feature self-interaction, then models the sensitivity of non-treatment features to specific treatments using a treatment-aware interaction module, and enhances the model's robustness through an intervention constraint module. M$^{3}$TN \cite{sun2024m} considers multi-treatment scenarios and uses a multi-task gating network to obtain uplift under different treatments. The latter focuses on modeling the causal effects of specific treatment. DESCN \cite{descn} designs an end-to-end deep cross network to comprehensively capture the information of intervention propensity, response and hidden intervention effects through multi-task learning. CDUM \cite{cdum} expands the treatment representation for guidance and indicator within a multi-task framework and proposes an online real-time feature learning network based on the characteristics of the video recommendation scenario.

However, the above methods either overemphasize the collaborative effects or the causal effects, making it difficult to adapt to short video recommendation scenarios which involves heterogeneous multi-treatment. Moreover, the short video recommendation scenario involves complex multi-responses, necessitating personalized and nuanced decision-making regarding multiple responses, which is also missing in current methods. Our proposed HMUM introduce a solution for the video recommendation scenario that can effectively capture both the synergistic effects between treatments and the individual causal effects of each treatment, and further simulates the joint benefits of multiple responses for online personalized adjustments.

%% file: sections/preliminary.tex
\section{PRELIMINARIES}
Before illustrating into the details of our proposed model, we break down the trade-off optimization problem in short-video recommendation into two components: the first is offline uplift modeling for multiple treatments, and the second is online decision-making. Below are the definitions of these problems.

\subsection{Definition of Uplift Modeling Problem}
\label{Offline Problem Definition}
As indicated by the characteristics of uplift modeling, a single user cannot be exposed to multiple treatments at the same time. Therefore, we first conduct a Randomized Controlled Trial (RCT) in a real-world setting to collect user response data under different treatments. We denote these samples set as $\mathcal{Z}=\{\mathbf{z}_1, \mathbf{z}_2, ..., \mathbf{z}_N\}$, where $\mathbf{z}_i = (\mathbf{x}_i, \mathbf{t}_i, \mathcal{Y})$ and $N$ is the number of all instances. $\mathbf{x}_i$ represents the user features of the $i$-th instance. While $\mathbf{t}_i$ represents the specific treatment features for the $i$-th instance. And it is worth noticing that the type of treatment is determined for a particular sample. Besides, $\mathcal{Y}=\{\mathbf{y}_{i1}, \mathbf{y}_{i2}, ..., \mathbf{y}_{iR}\}$ is the set of responses that we're focused on. $R$ is the number of all types of recorded responses. 

Following the Neyman-Rubin potential outcome framework \cite{neyman_rubin}, we denote $y_{ir}^{t_i}$ as the $r$-th type of response when the user in the $i$-th instance receives a specific treatment $t_i \in [1,K]$, and $y_{ir}^{t_0}$ as the response when the user is not treated.

As we can not observe $y_{ir}^{t_i}$ and $y_{ir}^{t_0}$ at the same time, there is no true uplift result for each instance, which is known as a key reason uplift modeling differs from traditional supervised learning. Thus, we need uplift modeling to estimate the expected set of individual treatment effect $\{\tau_r^k(x_i)\}_{r\in [1,R]}$ for each instance. Specifically, we have $\tau_r^k(x_i) = \mathbb{E}(y_{ir}^k|t_i=k,x_i)-\mathbb{E}(y_{ir}^0|t_i=0,x_i)$.

Different from the existing uplift methods, for each $k \in [1,K]$, we design different branches to estimate $\tau_r^k(x_i)$ individually (for each branch, we can get a pair of outcomes for treatment and non-treatment), modeling the accurate causal effects for different treatments. 
Besides, to capture the synergistic effects between treatment, we utilize a multi-task learning paradigm to jointly optimize our model and use KL divergence to constrain the representations of responses induced by non-treatment across branches. Thus, we can eventually get the set of pairs of outcomes from each branch, i.e., $\{(\mathbb{E}(y_{ir}^k|t_i=k,x_i),\mathbb{E}(y_{ir}^{0,k}|t_i=0,x_i))\}_{k \in [1,K]}$ to facilitate the final decision-making part.

\subsection{Definition of Decision-making Problem}
\label{Online Problem Definition}

After obtaining the set of pairs of outcomes from each branch, it is crucial to design a criterion to make full use of the estimated outcomes to make decision. First and foremost, we calculate the average of $\mathbb{E}(y_{ir}^{0,k}|t_i=0,x_i)$ from different branches to obtain the final adjusted estimates, and get $\mathbb{E}(y_{ir}^{0,*}|t_i=0,x_i)$. Besides, to balance each response, we calculate the relative estimating uplift with $\delta_r^k(x_i) = \mathbb{E}(y_{ir}^k|t_i=k,x_i)/\mathbb{E}(y_{ir}^{0,*}|t_i=0,x_i)-1$. 

Considering the trade-offs among different responses, we have developed a multi-task learning framework to estimate the value weights for various responses for each user. Specifically, we utilize real-time contextual features (i.e, video features of several recent user interactions) and request-level candidates (i.e, all candidate video features under a single user request) of users as inputs $\mathbf{s}_i$ and the proportion of different types of candidate samples as labels to train the model. The trained model can then predict the user's value weight at the request-level in real-time. Consequently, we obtain the final balanced comprehensive score $\phi_i^k = \sum_{r=1}^R\left(w_{ir}*\delta_r^k(x_i)\right)$, where $w_{ir}$ represents the value weights for the $r$-th response. For the final decision, we can simply apply the criterion $\phi_i^k > \sigma$ ($\sigma$ is the threshold value) to determine whether to enable or enhance the $k$-th treatment.

%% file: sections/method.tex
\section{METHOD}
\begin{figure*}[t]
	\centering
	\setlength{\belowcaptionskip}{-0.0cm}
	\setlength{\abovecaptionskip}{-0.0cm}
	\includegraphics[width=1.0\textwidth]{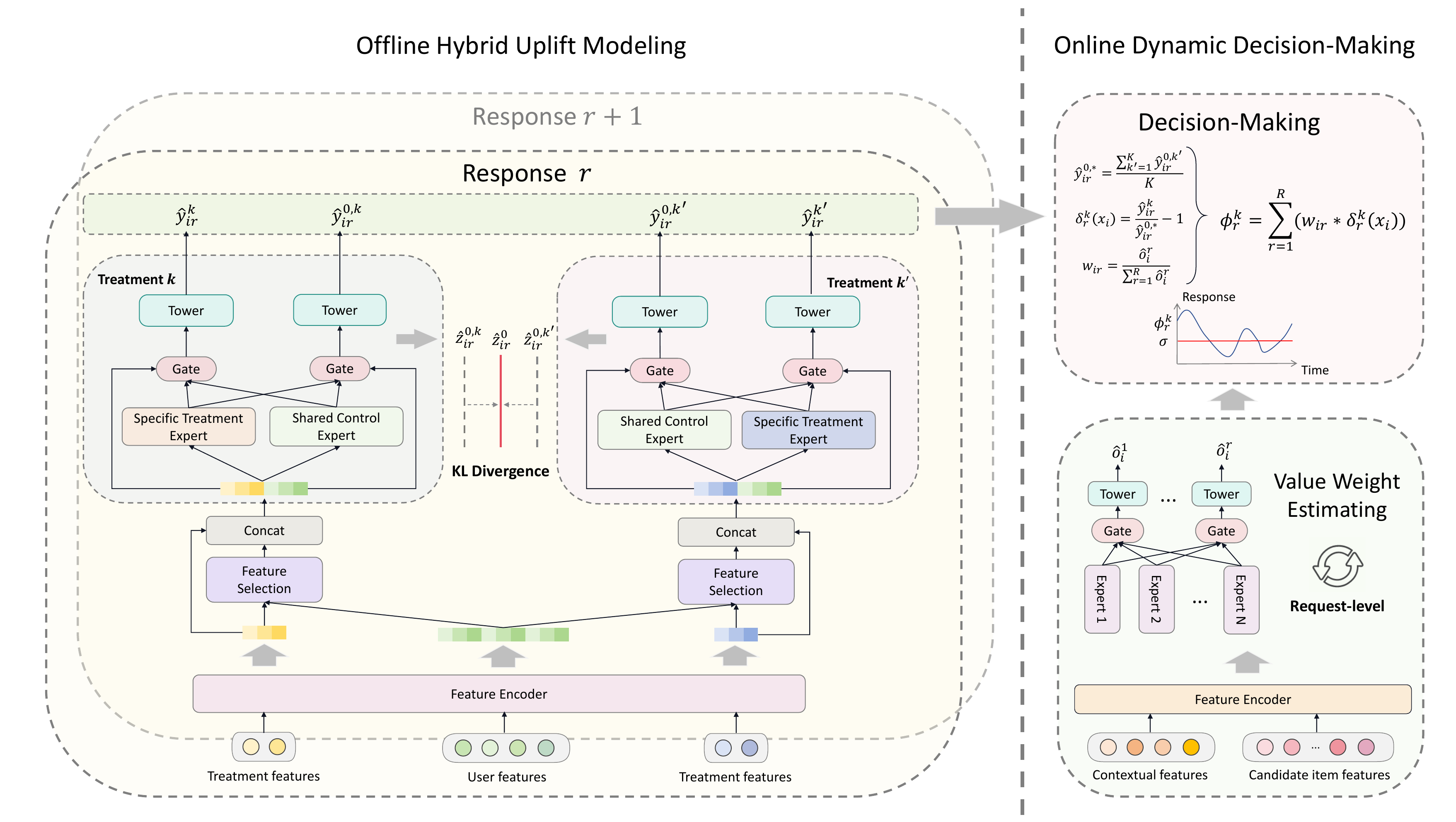}
	\caption{Illustration of the proposed HMUM framework. For the convenience of demonstration, the offline part of the model takes two responses and two treatments as an example.}
	\label{fig:framework}
	\vspace{-3mm}
\end{figure*}

We propose Heterogeneous Multi-treatment Uplift Modeling for Trade-off Optimization in Short-Video Recommendation (HMUM), which includes two modules: Offline Hybrid Uplift Modeling (HUM) and Online Dynamic Decision-Making (DDM). Figure \ref{fig:framework} illustrates the technical details of the proposed model.

\subsection{Offline Hybrid Uplift Modeling}
In this section, we introduce a module called Offline Hybrid Uplift Modeling.  This module features a tree structure that integrates individual causal effect branches with joint optimization for multiple treatments. For simplicity and without loss of generality, we will focus on a specific user response (the $r$-th response), denoted as $y_{ir}$ for the $i$-th instance, to illustrate our proposed modules.

\subsubsection{Feature Selection Module.}
For each instance $\mathbf{z}_i = (\mathbf{x}_i,\mathbf{t}_i,y_{ir})$, we encode both $\mathbf{x}_i$ and $\mathbf{t}_i$ into embeddings. We begin by discretizing the continuous numerical features. Following this, we apply an Embedding Look-up operation to transform the original features into embeddings. Specifically, we proceed as follows:
\begin{equation}
\mathbf{e}_{xi} = \mathbf{E}_x^{T} \cdot \mathbf{x}_i, \ 
\mathbf{e}_{ti} = \mathbf{E}_t^{T} \cdot \mathbf{t}_i,
\label{embedding convert}
\end{equation}
where $\mathbf{E}_x \in \mathbb{R}^{f_x \times d}$ and $\mathbf{E}_t \in \mathbb{R}^{f_t \times d}$ are the created embedding tables, $f_x$ and $f_t$ are the numbers of non-treatment features and treatment features, and $d$ is the embedding size. 

Furthermore, we utilize a target attention mechanism to extract the relevant features from the non-treatment embeddings with the help of specific treatment embeddings. And have:
\begin{equation}
\left\{\begin{array}{c}
\begin{aligned}
\mathbf{w}_i^{k} &= Softmax(\mathbf{W}^k \cdot \mathbf{e}_{ti} + \mathbf{b}^k) \\
\mathbf{e}_{xi}^{k} &= \sum_{j=1}^{f_x} \left({\mathbf{w}_i^{k}(j) \cdot \mathbf{e}_{xi,j}}\right)
\end{aligned}
\end{array}\right.
\end{equation}
where $\mathbf{e}_{xi}^{k}$ denotes the extracted non-treatment embedding for the $k$-th (corresponding to $t_i$) treatment. $\mathbf{W}^k \in \mathbb{R}^{f_x \times d}$ and $\mathbf{b}^k \in \mathbb{R}^{f_x \times 1}$ are feature transformation matrix and bias matrix. And $\mathbf{w}_i^{k} \in \mathbb{R}^{f_x \times 1}$ is the attention vector which is used as selector to calculate the weighted sum of all non-treatment embeddings. $\mathbf{w}_{i}^{k}(j)$ denotes the $j$-th element of vector $\mathbf{w}_i^{k}$ and $\mathbf{e}_{xi,j}$ is the $j$-th non-treatment feature. It is worth noting that in the case of $k=0$, the tree structure we designed follows a multi-branch joint optimization approach. Therefore, the parameters $\mathbf{W}^k$ and $\mathbf{b}^k$ under the corresponding branch are shared with $k$ under specific values.

Additionally, we use a concatenate operation to combine the extracted non-treatment embedding with the treatment embedding together as the inputs to the next part:
\begin{equation}
\mathbf{f}_{i}^{k} = \mathop{Concat}(\mathbf{e}_{xi}^{k},\mathbf{e}_{ti}).
\end{equation}

\subsubsection{Individual Causal Effect Branches.}
To efficiently model the individual causal effects of each treatment, we design a branching structure where each branch specifically takes its corresponding $\mathbf{f}_{i}^{k}$ (where $k \in [1,K]$) as input. This structure outputs the predicted values $\mathbb{E}(y_{ir}^k|t_i=k,x_i)$ and $\mathbb{E}(y_{ir}^{0,k}|t_i=0,x_i)$, which represent the individually estimated responses for the corresponding treatment. 

Specifically, take the $k$-th branch as example, we use operations such as multilayer perceptron to generate $M$ expert representations (denoted as $\mathbf{p}_m^k$, $m\in\{1,2,...,M\}$) from $\mathbf{f}_{i}^{k}$. Following the classical operations of multi-task learning \cite{meng2023parallel,PLE,cigf,copf,tang2024touch,weng2023curriculum}, we employ a gating mechanism to aggregate the expert information. The final score is then output through a specific tower, using this accumulated information as input. This process can be formulated as:
\begin{equation}
\left\{\begin{array}{c}
\begin{aligned}
    \mathbf{g}_i^{k} &= Softmax(\mathbf{W}_g^k \cdot \mathbf{f}_{i}^{k} + \mathbf{b}_g^k) \\
    \hat{y}_{ir}^{k} &= h^{k}\left(\sum_{m=1}^{M} \left({\mathbf{g}_i^{k}(m) \cdot \mathbf{p}_m^k}\right)\right)
\end{aligned}
\end{array}\right.
\end{equation}
where $\mathbf{z}_i^k$ is the output of corresponding gate. $\mathbf{W}_g^k \in \mathbb{R}^{M \times d}$ and $\mathbf{b}_g^k \in \mathbb{R}^{M \times 1}$ are feature transformation matrix and bias matrix, and $\mathbf{g}_i^{k} \in \mathbb{R}^{M \times 1}$ is the attention vector which is used as selector to calculate the weighted sum of all experts. $\mathbf{g}_{i}^{k}(m)$ denotes the $m$-th element of vector $\mathbf{g}_i^{k}$. $h^{k}(\cdot)$ are the tower functions.

Furthermore, in the case of $k=0$, the process is similar across branches. However, each branch produces distinct gate outputs and predicted values at this stage. To provide clarity, for a specific branch $k'$, we denote $\mathbf{z}_{ir}^{0,k'}$ ($k' \neq 0$) and $\hat{y}_{ir}^{0,k'}$ ($k' \neq 0$) as the individual gate output and predicted value for the $k=0$ scenario in that branch.

\subsubsection{Multi-treatment Joint Optimization and Prediction.}
As we have obtained $\hat{y}_{ir}^{k}$ and $\hat{y}_{ir}^{0,k'}$ from all branches, we further use the mean square error as the main training loss function. Besides, to constrain the representations induced by non-treatment across branches, we apply KL divergence to fix the original loss. This can be formulated as:
\begin{equation}
\mathcal{L} =
\left\{\begin{array}{c}
\begin{aligned}
 &\frac{1}{N} \sum_{i=1}^{N}\sum_{k'=1}^{K}\left((y_{ir}^{k} - \hat{y}_{ir}^{0,k'})^2+ D_{\text{KL}}(\mathbf{z}_{ir}^{0,k'} \parallel \frac{\sum_{k'=1}^K\mathbf{z}_{ir}^{0,k'}}{K})\right),  &k = 0 \\
&\frac{1}{N} \sum_{i=1}^{N} (y_{ir}^{k} - \hat{y}_{ir}^{k})^2,  &k \neq 0
\end{aligned}
\end{array}\right.
\end{equation}
where $y_{ir}^{k}$ is the response label under $k$-th treatment. It is worth noting that in the training stage, since each sample $(\mathbf{x}_i,\mathbf{t}_i,y_{ir})$ is obtained by recording response only under the action of a specific treatment (e.g, $k$-th treatment), we only use the output $\hat{y}_{ir}^{k}$ ($k \neq 0$) and $\hat{y}_{ir}^{0,k'}$ ($k=0, k' \in [1,K]$) for training, while mask operation is applied to the calculation of loss to other towers.

Finally, at the prediction stage, we can infer $\hat{y}_{ir}^{k}$ and $\hat{y}_{ir}^{0,k'}$ for each user by replacing the feature of treatment $\mathbf{t}_i$ iteratively. Further, we can obtain the estimated response scores of users for different treatments, and deliver them to the next online part, DDM.

\subsection{Online Dynamic Decision-Making}
As briefly summarized in Section \ref{Online Problem Definition}, we will now provide the detailed implementation of DDM as a supplement for better understanding.

\subsubsection{Pre-processing of Estimates.}
In order to achieve a more accurate estimate of user response, we consider the following constraint:
\begin{equation}
   \hat{y}_{ir}^{0,min} \leq \frac{\sum_{k'=1}^K \hat{y}_{ir}^{0,k'}}{K} \leq \hat{y}_{ir}^{0,max}
\end{equation}
where $r$ denotes the $r$-th type of responses. $K$ is the number of branches. $\hat{y}_{ir}^{0,min}$ and $\hat{y}_{ir}^{0,max}$ are the minimum and maximum values of $\hat{y}_{ir}^{0,k'}$, respectively, $\forall k' \in [1,K]$. 

Therefore, we use the average value $\frac{\sum_{k'=1}^K \hat{y}_{ir}^{0,k'}}{K}$ as the adjusted approximate value for the final estimated response without treatment. We denote this value as $\hat{y}_{ir}^{0,*}$.

Additionally, to bridge the gap between the different response distributions, we use a new score to represent the relative uplift:
\begin{equation}
    \delta_r^k(x_i) = \frac{\hat{y}_{ir}^{k}}{\hat{y}_{ir}^{0,*}}-1.
\end{equation}

\subsubsection{Value Weight Estimating.}
\label{online train}

Taking into account the trade-offs among different responses, we have developed a multi-task learning framework to estimate the value weights of various responses for each individual user. Specifically, we use contextual features and request-level candidate item features as inputs, i.e., $\mathbf{s}_i$. At the same time, we use the proportions of different types of exposed samples as labels.

We first encode the online features to embeddings. This process is similar to Formula (\ref{embedding convert}), thus we get the online embeddings $\mathbf{e}_{si} = \{\mathbf{e}_{si}^0,\mathbf{e}_{si}^1,...,\mathbf{e}_{si}^{f_s}\}$. Here $\mathbf{e}_{si}^j \in \mathbb{R}^{1 \times L^j \times d_s}, j \in [0,f_s]$, $L^j$ represents the number of candidates in a request-level of users' historical datas, $d_s$ is the embedding size. Then, we apply an average pooling to the second dimension, and concatenate all the pooled embeddings together. This process can be represented as:
\begin{equation}
    \mathbf{e}_{si}^* = \mathop{Concat}\limits_{j=0}^{f_s}\left(\sum\limits_{l=1}^{L^j}{\mathbf{e}_{si}^j(l)/(L^j)}\right),
\end{equation}
where $\mathbf{e}_{si}^j(l) \in \mathbb{R}^{1\times1\times d_s}$ denotes the $l$-th interacted item embedding.

Further, we equip $\mathbf{e}_{si}^*$ with a multi-task learning network. Here, the form of the network is similar to MMOE \cite{mmoe}, and for simplify, we denote $\hat{o}_i^r$ ($r \in [1,R]$) as the output of each tower of this part. Here, $R$ is the number of all types of exposed samples, which is corresponding to the number of types of recorded user responses.

For the training details, we design an online request-level label based on the proportions of different types of exposed samples. Specifically, using consumption-oriented responses and commercial responses as examples of two response types, let $V_1$ represents the number of exposed videos where the consumption score is higher than the commercial score (the conclusion is drawn by comparing the ranking percentile of a video's commercial metric and consumption metric within the dataset), and $V_2$ represents the number of videos where the opposite is true. Without loss of generality, we have:
\begin{equation}
    o_i^r = \frac{V_r}{V}, \sum_{r=1}^R V_r = V
\end{equation}
where $o_i^r$ is the label. $V_r$ denotes the number of exposed samples of the $r$-th type (where the $r$-th type has the highest score), and $V$ is the total number of exposed samples in the user's request.

Additionally, we utilize the mean square error as loss function to optimize the model. While for the prediction, we deploy this module directly online, and it is invoked with every user request to calculate the real-time interest score $\hat{o}_i^r$.

Last but not least, we calculate the value weight with a common proportional formula:
\begin{equation}
    w_{ir} = \frac{\hat{o}_i^r}{\sum_{r=1}^R \hat{o}_i^r}
\end{equation}
where $w_{ir}$ represents the value weights for the $r$-th response.

\subsubsection{Decision-making for Multiple Responses.}

As a result, we derive the final balanced comprehensive score:
\begin{equation}
    \phi_i^k = \sum_{r=1}^R\left(w_{ir}*\delta_r^k(x_i)\right)
\end{equation}
where $\delta_r^k(x_i)$ indicates the corresponding contribution for the $k$-th treatment.

For the final decision-making process, we apply the criterion $\phi_i^k > \sigma$ (where $\sigma$ is a predefined threshold)to determine whether to activate or enhance the $k$-th treatment.

This allows us to dynamically and personally balance trade-offs, enabling specific treatments for each user, which enhances revenue while improving the user experience.

%% file: sections/experiments.tex
\section{OFFLINE EVALUATIONS}

\subsection{Experimental Setting}

\subsubsection{Dataset Description}



We assess our proposed method using both public and real-world industrial datasets. For the public datasets, we utilize CRITEO \cite{criteo} and LAZADA \cite{descn}. The CRITEO dataset is based on a large-scale advertising scenario, which was created by Criteo AI Labs. It includes nearly 14 million instances, 12 features, a treatment indicator and two labels (i.e., visit and conversion). We use "visit" as the target label and randomly split the dataset into training, validation, and testing sets with a ratio of 8:1:1. The LAZADA dataset comes from a real-world voucher distribution scenario on the Lazada e-commerce platform. It contains 83 features, a treatment indicator and a label. Due to the operational target strategy, treatment allocation is selective in the actual production environment. As such, the data with treatment bias is used as the training set, while a small number of users not affected by the targeting strategy are used as the testing set, where their treatment allocation follows a randomized controlled trial. 

The industrial dataset comprises online user feature data collected 14 days before the initiation of multiple randomized controlled trial (RCT) experiments, which include various treatments and control groups.  It also contains user feedback collected during these RCT experiments, specifically focusing on app usage time and video views.  The dataset features encompass users' inherent attributes (such as region and activity level), consumption information from the previous 7 and 30 days (including usage duration and video views), and commercial attributes (like user commercial value classification).  We used the average app usage time and video view counts during the experiment as labels. In addition, we define the scorer that assigns the consumption-side score as treatment 1 ($t_1$) and the scorer that assigns the commercialization-side score as treatment 2 ($t_2$), and through a posterior analysis to ensure that it satisfies heterogeneity. The dataset includes approximately 6,000,000 instances, which we divided into training and test sets with an 8:2 ratio.

\subsubsection{Evaluation Metrics}
In our experiments, we adopt two widely used metrics to evaluate uplift ranking performance of different models: normalized Area Under the Uplift Curve (AUUC) and normalized area under the qini curve (QINI). We refer to the implementation of the Python package scikit-uplift\footnote{https://www.uplift-modeling.com/en/latest} to calculate these metrics. For experiments on public datasets, we use standard Python package; For experiments on industrial datasets, we modify the calculation method based on the characteristics of continuous labels. To be specific, we normalize the labels before calculating the metrics, and adjust the calculation method of the perfect uplift curve in the source code accordingly.

\subsubsection{Baseline Models}
To demonstrate the effectiveness of our proposed HMUM, we selected a set of currently popular and representative uplift modeling methods, including:  S-Learner \cite{metalearners}, T-Learner \cite{metalearners}, BNN \cite{bnn}, TARNet \cite{shalit2017estimating}, CFRNet \cite{shalit2017estimating}, CEVAE\cite{cevae}, GANITE \cite{ganite}, DragonNet \cite{dragonnet}, FlexTENet \cite{flextenet}, SNet \cite{snet}, EUEN \cite{euen}, DESCN \cite{descn}, and EFIN \cite{efin}.

For the experiments on industrial dataset, we employ some multi-treatment models in order to test the effectiveness, including EFIN and M$^{3}$TN \cite{sun2024m}. In addition, we extend S-Learner, T-Learner and TARNet to adapt them to multi-treatment tasks. Specifically, for S-Learner, we take multi-treatment features directly as input. For T-Learner and TARNet, we arrange independent branches to calculate ITE for each treatment.


\subsection{Performance Comparison}

\begin{table*}[ht]
\setlength{\abovecaptionskip}{0cm}
\setlength{\belowcaptionskip}{0mm}
\caption{The overall performance comparison on industrial offline dataset. Boldface denotes the highest score and underline indicates the results of the best baselines.}
    \centering
    \begin{threeparttable}
	\resizebox{\linewidth}{!}{
    \begin{tabular}{c|cc|cc|cc|cc}
    \toprule
    {Response}&\multicolumn{4}{c}{APP usage time}&\multicolumn{4}{|c}{Video view counts}\cr
    \cmidrule(lr){1-1} \cmidrule(lr){2-5} \cmidrule(lr){6-9}
    {Treatment}&\multicolumn{2}{c}{Treatment 1}&\multicolumn{2}{|c}{Treatment 2}
    &\multicolumn{2}{c}{Treatment 1}&\multicolumn{2}{|c}{Treatment 2}\cr
    \cmidrule(lr){1-1}
    \cmidrule(lr){2-3} \cmidrule(lr){4-5} \cmidrule(lr){6-7}  \cmidrule(lr){8-9} 
    {Metrics} &QINI($10^{-4}$)&AUUC($10^{-4}$)&QINI($10^{-4}$)&AUUC($10^{-4}$)&QINI($10^{-4}$)&AUUC($10^{-4}$)&QINI($10^{-4}$)&AUUC($10^{-4}$) \cr
    \midrule
    S-Learner &8.81	&11.91	&-11.01	&-14.92	&\underline{13.37}	&\underline{15.95}	&10.28	&12.23 \cr
    T-Learner &8.69	&11.76	&-22.14	&-30.03	&5.62	&6.72	&10.88	&13.02 \cr
    TARNet &7.07	&9.56	&\underline{13.18}	&\underline{17.89}	&-0.24	&-0.29	&7.61	&9.09 \cr
    EFIN &10.68	&14.47	&-5.3	&-7.21	&-8.07	&-9.59	&10.97	&13.07 \cr
    M$^{3}$TN &\underline{13.04}	&\underline{17.66}	&-17.74	&-24.08	&1.25	&1.48	&\underline{24.4}	&\underline{29.18} \cr
    \hline 
    HUM w/o KL &10.87	&14.71	&14.59	&19.78	&5.00	&5.97	&15.56	&18.63 \cr
    \textbf{HUM} &\textbf{14.55}	&\textbf{19.67}&\textbf{19.90}&\textbf{27.01}&\textbf{15.87}&\textbf{18.95}&\textbf{25.44}&\textbf{30.46}\cr
    \bottomrule
    \end{tabular}
    }
    \end{threeparttable}
    \label{comparisons_model222}
    \vspace{-3mm}
\end{table*}


\begin{table}[ht]
   \setlength{\abovecaptionskip}{-0.0cm}
   \setlength{\belowcaptionskip}{-0.0cm}
\caption{The overall performance comparison on Lazada and Criteo datasets. Boldface denotes the highest score and underline indicates the results of the best baselines.}
    \centering
    \begin{threeparttable}
	\resizebox{0.85\linewidth}{!}{
    \begin{tabular}{c|cc|cc}
    \toprule
    {Dataset}&
    \multicolumn{2}{c}{LAZADA}&\multicolumn{2}{|c}{CRITEO}\cr
    \cmidrule(lr){1-1}
    \cmidrule(lr){2-3} \cmidrule(lr){4-5} 
    {Metrics}&QINI&AUUC&QINI&AUUC\cr
    \midrule
    S-Learner &0.0207&0.0029&0.0902 & 0.0354  \cr
    T-Learner &0.0217&0.0031&0.0790 & 0.0310 \cr
    BNN &0.0230&0.0033&0.0888 & 0.0349  \cr
    TARNet &0.0214&0.0030&0.0847 & 0.0333 \cr
    CFRNet &0.0231&0.0033&0.0901 & 0.0354 \cr
    CEVAE &0.0158&0.0022&0.0867 & 0.0341  \cr
    GANITE &0.0174&0.0024&0.0818 & 0.0322  \cr
    DragonNet &0.0176&0.0025&0.0851 & 0.0335  \cr
    FlexTENet &0.0185&0.0026&\underline{0.0924} & \underline{0.0363}  \cr
    SNet &\underline{0.0238}&\underline{0.0034}&0.0843 & 0.0331  \cr
    EUEN &0.0156&0.0022&0.0898 & 0.0353  \cr
    DESCN &0.0223&0.0031&0.0803 & 0.0316  \cr
    EFIN &0.0141&0.0020&0.0859 & 0.0337  \cr
    \hline    \textbf{HUM}&\textbf{0.0278}&\textbf{0.0039}&\textbf{0.0933}&\textbf{0.0366}\cr
    \bottomrule
    \end{tabular}}
    \end{threeparttable}
    \label{comparisons_model111}
    \vspace{-3mm}
\end{table}

\subsubsection{Performance comparison under heterogeneous multi-treatment}
Table \ref{comparisons_model222} presents the evaluation results of the offline module HUM in our proposed HMUM and baselines suitable for multiple treatments on the industrial dataset that contains heterogeneous multi-treatments. To ensure a fair comparison, the implementation of the baseline models also follows the aforementioned parameter settings. In addition, all baselines and our proposed model use the same initialization embedding method and no hyperparameter adjustment is performed during training. We can see that:

\begin{itemize} 
    \item In video recommendation scenarios, the heterogeneity of multi-treatments is common, and the effects of different treatments on the same response are often mutually exclusive.  Inappropriate using multi-treatment uplift modeling can lead to suboptimal overall performance and even negative uplift (i.e., the metric with a value less than 0 in the table), which can be described as "downlift".
    \item The multi-treatment uplift baseline model shows significant differences in modeling effectiveness between the two treatments, which may be due to the lack of individual modeling.
    \item Our proposed model achieves the best performance in uplift modeling for different user responses across all treatments, demonstrating its superiority and advanced capabilities.
\end{itemize}

\subsubsection{Performance comparison under a single treatment}
To further explore the broad applicability of our model, we also conduct experiments on single-treatment public datasets, Lazada and Criteo. Inspired by multi-interest modeling, we treat different branches in HUM as different preference modeling branches under the same treatment for testing. According to the results shown in Table \ref{comparisons_model111}, different baselines exhibit certain performance variations when evaluated on the two public datasets, failing to provide consistent and reliable improvements across different scenarios. In contrast, our model HUM also achieves state-of-the-art performance on the offline single-treatment datasets. This may be due to the 
multi-branch structure of our model, which jointly models the single treatment from different angles, thereby eliminating bias and achieving such effects. This again demonstrates the wide applicability of our proposed HUM.


\subsubsection{Impact of the KL divergence}

To assess the effectiveness of the key design element of the offline module HUM in our proposed method HMUM, we created a HUM variant without the KL divergence constraint (HUM w/o KL). The results are shown in Table \ref{comparisons_model222}. It can be observed that removing the KL divergence leads to a decrease in performance, highlighting the importance of constraining the non-treatment under different branches. This demonstrates the effectiveness of our proposed module.

%% file: sections/application.tex
\section{ONLINE APPLICATION}
Our proposed HMUM can be deployed at various stages in the online environment, where different strategies or modules can be treated as distinct treatments or interventions. HMUM effectively models the causal effects of these multiple treatments, allowing for the dynamic selection of the most promising users. To evaluate the model's online performance in real-world scenarios, we integrate HMUM into a large-scale industrial recommendation pipeline serving hundreds of millions of users. We implement HMUM in two critical components of the online service: Ranking and Edge Rerank, and conduct comprehensive A/B testing in both components.

Specifically, we conduct a 13-day online test for the Ranking component. For the Edge Rerank component, we conducted a similar experimental setup, testing online over an 8-day period.

\begin{figure}[t]
	\centering
	\setlength{\belowcaptionskip}{-0.0cm}
	\setlength{\abovecaptionskip}{-0.0cm}
	\includegraphics[width=\linewidth]{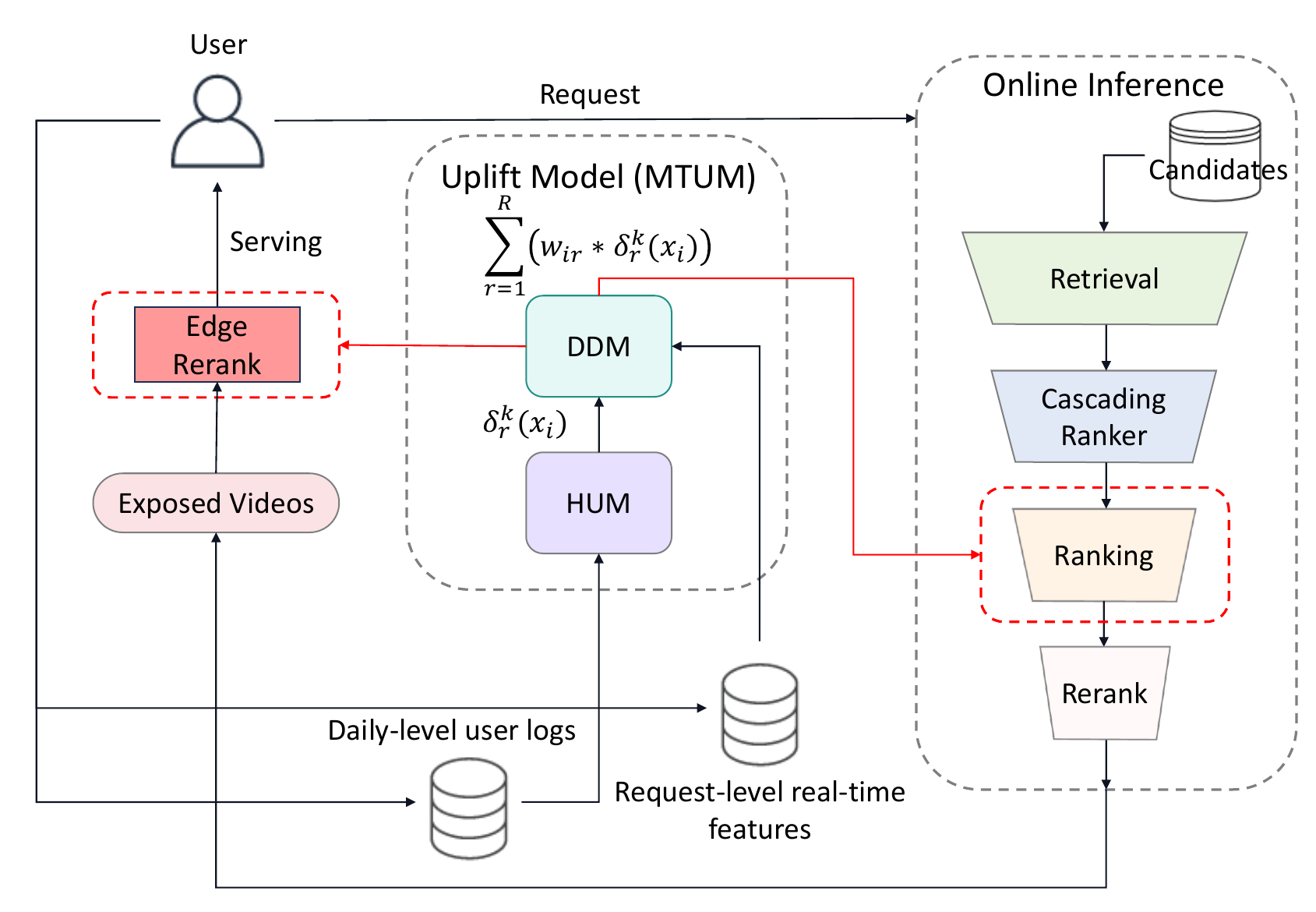}
	\caption{Illustration of the online pipeline.}
	\label{fig:online_pipeline}
	\vspace{-5mm}
\end{figure}

\subsection{System Description}

In the online video recommendation pipeline, shown in Figure \ref{fig:online_pipeline}, a sequence of modules works in tandem to filter candidate videos, ultimately presenting the selected videos to the user. When a user initiates a request, their attributes and contextual features are transmitted to the online service. This activates the pipeline, which progressively filters and selects videos from the candidate pool in a cascading manner. Last but not least, the chosen exposed videos are sent to the Edge Rerank which reorders the videos, and then shows them to user, completing the request process.

\subsection{Scheme Overview}
Our model, as illustrated in Figure \ref{fig:online_pipeline}, is deployed in several components of online pipeline (i.e., the server-side, represented by Ranking, and the client-side, represented by Edge Rerank \cite{gong2020edgerec,gong2022real} ). Meanwhile, We use APP usage time and Video view counts as labels. Our task is to use HMUM to jointly model these two treatments and identify the ideal users for each treatment to be effective. These scorers are then utilized to evaluate and rank the candidate videos, a process that is consistent across all stages of the online environment.

First, we use the trained HUM to infer the value of $\delta_r^k(x_i)$ for all users on the platform at a daily-level and store these values in a database system, such as Redis. Then, for each user request, we use DDM to perform real-time inference based on the user's request-level features, obtaining the user's value weight $w_{ir}$ for different responses. Finally, for each $k \in [1,K]$, we use the decision-making condition $\sum_{r=1}^R\left(w_{ir}*\delta_r^k(x_i)\right)>\sigma$ to determine the enablement of treatments at the level of the user's request.

\subsection{Online Experimental Results} 

\subsubsection{Online A/B test results.}
We utilize standard industry metrics for consumption (APP usage time), along with commercial metrics (Video view counts) to evaluate online performance. We conduct A/B tests at both the Ranking and Edge Rerank stages. As shown in Table \ref{online_ab_ranking}, we can draw the following conclusions:

\begin{itemize}
    \item Our proposed method achieved significant improvements in consumption metrics at both stages, while also achieving a slight positive increase in commercialization metrics. This effectively addresses the trade-off issue under multi-treatment conditions.
    \item The outstanding performance across multiple stages further underscores the broad compatibility and applicability of our approach.
\end{itemize}

As a result, our model is now fully deployed at multiple stages of the Kuaishou platform, serving hundreds of millions of users daily.

\begin{table}[t]
   \setlength{\abovecaptionskip}{0cm}
   \setlength{\belowcaptionskip}{-0.0cm}
\caption{The performance of Online A/B Testing at Ranking stage and Edge Rerank stage on the Kuaishou platform.}
    \centering
    \begin{threeparttable}
	\resizebox{0.9\linewidth}{!}{
    \begin{tabular}{c|c|c}
    \toprule
    {Stage}&
    {Ranking} &
    {Edge Rerank} \cr
    \cmidrule(lr){1-1}
    \cmidrule(lr){2-2} 
    \cmidrule(lr){3-3} 
    {Metrics}  &\textbf{Imprv.}  &\textbf{Imprv.}\cr
    \midrule
    APP usage time &\textbf{\color{red}{+0.044\%}} &\textbf{\color{red}{+0.073\%}}\cr
    APP usage time (per capita) &\textbf{\color{red}{+0.055\%}} &\textbf{\color{red}{+0.072\%}}\cr
    Video view counts  &\color{red}{+0.007\%}  &\color{red}{+0.002\%}\cr
    \bottomrule
    \end{tabular}
    }
    \end{threeparttable}
    \label{online_ab_ranking}
    \vspace{-3mm}
\end{table}


\subsubsection{Ablation Test of the Treatment.}

\begin{table}[t]
   \setlength{\abovecaptionskip}{0cm}
   \setlength{\belowcaptionskip}{-0.0cm}
\caption{The ablation experiments at Edge Rerank stage on the Kuaishou platform.}
    \centering
    \begin{threeparttable}
	\resizebox{\linewidth}{!}{
    \begin{tabular}{c|ccc}
    \toprule
    {Stage}&
    \multicolumn{3}{c}{Edge Rerank} \cr
    \cmidrule(lr){1-1}
    \cmidrule(lr){2-4}
    {Variants} & \text{Single $s_1$}& \text{Single $s_2$}& \text{Both $s_1$ \& $s_2$}\cr
    \midrule
    APP usage time &\textbf{\color{red}{+0.125\%}}&\textbf{\color{lightgreen}{-0.094\%}}&\textbf{\color{lightgreen}{-0.044\%}}\cr
    APP usage time (per capita) &\textbf{\color{red}{+0.085\%}}&\textbf{\color{lightgreen}{-0.078\%}}&\textbf{\color{lightgreen}{-0.056\%}}\cr
    Video view counts &\textbf{\color{lightgreen}{-0.227\%}}&\textbf{\color{red}{+0.151\%}}&\color{red}{+0.091\%}\cr
    \bottomrule
    \end{tabular}}
    \end{threeparttable}
    \label{online_ab_ablation}
    \vspace{-3mm}
\end{table}

In this section, we conduct ablation experiments to further validate the effectiveness of our method. Specifically, we design three variants to test, i.e., (1) Single $s_1$: we implement the consumption-side score in the form of a scorer across the entire population; (2) Single $s_2$: we implement the commercialization-side score in the form of a scorer across the entire population; (3) Both $s_1$ \& $s_2$: we implement both strategy 1 and strategy 2 with a form of multiplicative scorer across the entire population. Following these applications, we make parameter adjustments to optimize performance. This process highlights the trade-off issues that can occur when implementing either a single strategy or a combined strategy for the \emph{entire population}.

As illustrated in Table \ref{online_ab_ablation}, we can conclude the following:
\begin{itemize}
    \item These strategy variants applied to the entire population all lead to significant trade-off issues.
    \item In contrast to the results in Table \ref{online_ab_ranking}, our HMUM model performs uplift modeling for each strategy and dynamically applies specific strategies to each user based on their real-time features. This approach successfully resolves the trade-off issues, further demonstrating the effectiveness of our method.
\end{itemize}

%% file: sections/conclusion.tex
\section{CONCLUSIONS AND FUTURE WORK}
In this work, we have addressed the complex challenges of optimizing short-video recommendation systems through the development of the Heterogeneous Multi-treatment Uplift Modeling (HMUM) framework. By tackling the intricacies of multiple treatment effects and personalized decision-making, HMUM effectively balances diverse user responses such as APP usage time and video view counts. The Offline Hybrid Uplift Modeling (HUM) module captures both individual and synergistic treatment effects, while the Online Dynamic Decision-Making (DDM) module dynamically adjusts response value weights based on user characteristics. Our extensive evaluation on public datasets and the industrial platform demonstrates that HMUM significantly enhances recommendation effectiveness, achieving notable improvements in key consumption metrics. The successful deployment of HMUM underscores its practical applicability and impact on real-world systems. 

For future work, we plan to apply our approach to other stages of the video recommendation pipeline and enhance the applicability of the model in a broader range of scenarios, such as live streaming recommendation. In addition, we are also interested in exploring more effective ways to utilize heterogeneous multi-treatment and provide valuable insights to the community, fostering further exploration and innovation in uplift modeling for video recommendation scenarios.

%% file: sections/appendix.tex
\section{APPENDIX}
\label{APPENDIX}
\subsection{Parameter Settings}
Our proposed HMUM is implemented in TensorFlow \cite{TensorFlow}. We use the Adam optimizer \cite{adam} with a learning rate of 0.001 and set the learning rate reduction factor to 0.6. The number of allowed epochs with no improvement is 2, after which the learning rate will be reduced. In addition, we set the batch size to 4096 and the embedding dimension to 32. The default value of the threshold $\sigma$ is 0. 



\subsection{Indepth Analysis}


\begin{figure}[ht]
        \centering
        \setlength{\belowcaptionskip}{0cm}
	\setlength{\abovecaptionskip}{0cm}
	\subfigure{
        \begin{minipage}[t]{0.47\linewidth}
        \centering
		\label{fig:coefficient_beibei} 
		\includegraphics[width=\textwidth]{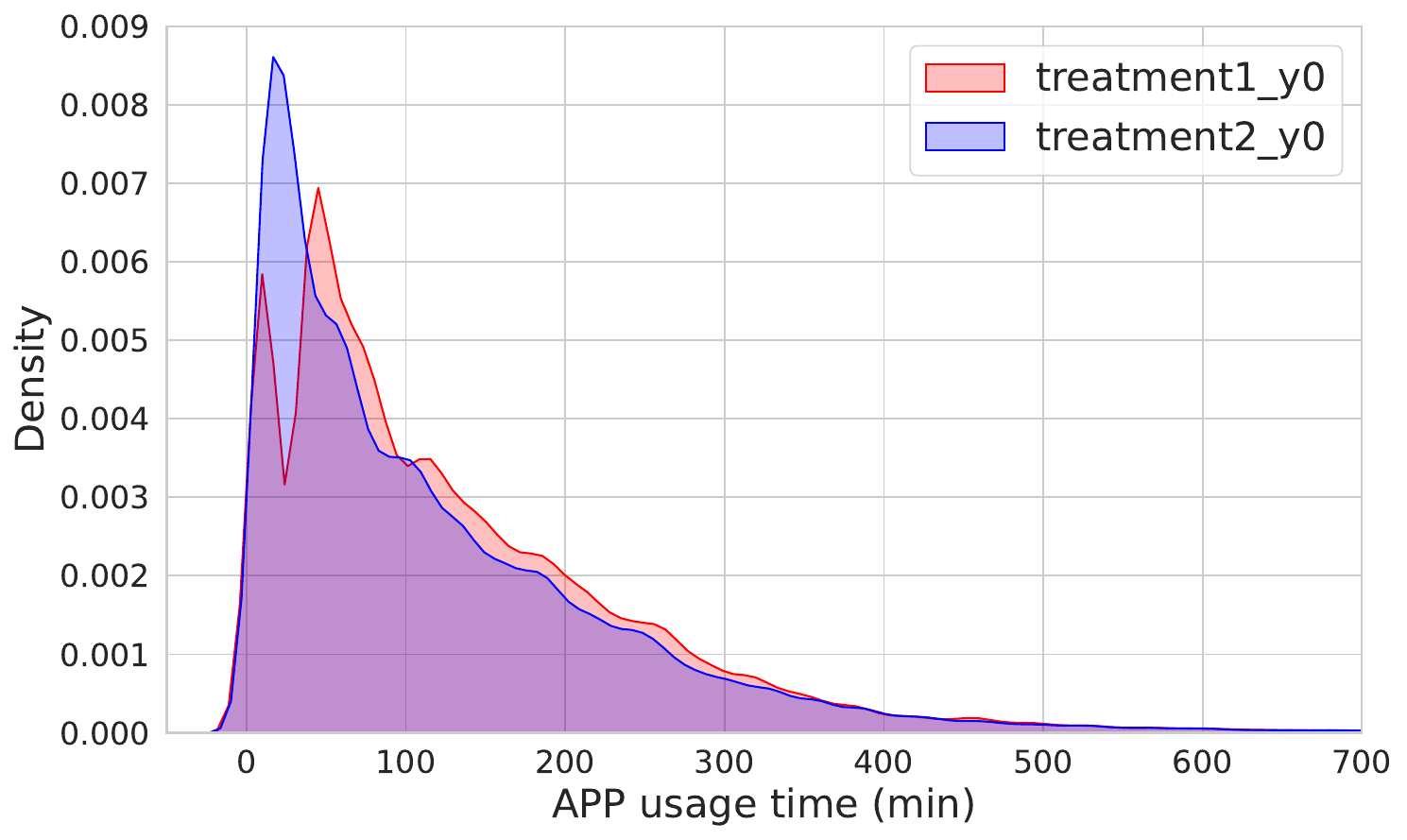}
        \end{minipage}}
	\subfigure{
        \begin{minipage}[t]{0.47\linewidth}
        \centering
		\label{fig:coefficient_taobao} 
		\includegraphics[width=\textwidth]{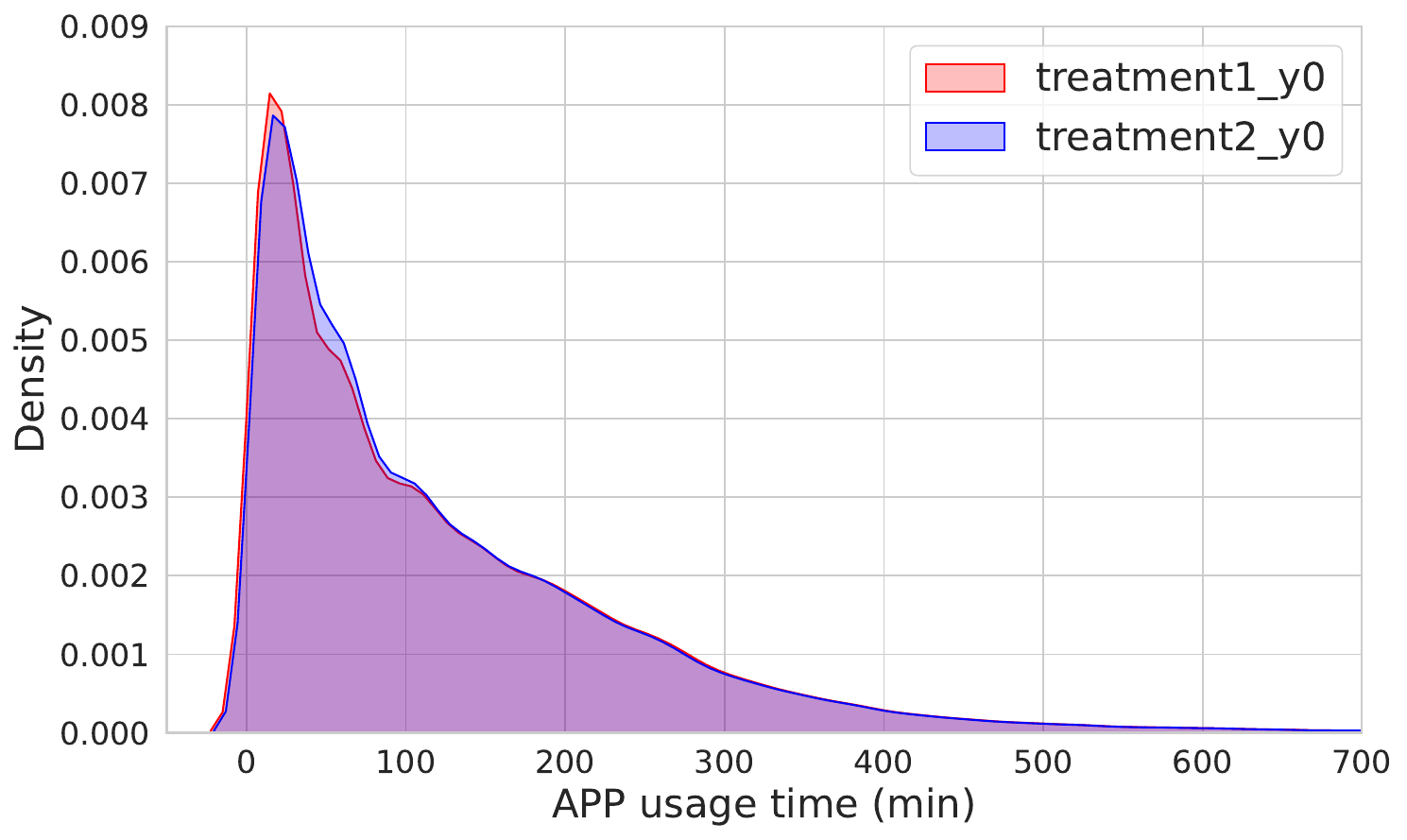}
        \end{minipage}}
	\caption{Comparison of the non-treatment output distributions across different branches under APP usage time. The left figure shows independent modeling of each treatment, while the right figure shows joint modeling of each treatment (i.e., HMUM).}
	\label{fig:compare_app}
    \vspace{-5mm}
\end{figure}

We utilize single-sided framework to independently model the uplift of different treatments across each response in industrial dataset and compare it with HMUM. Due to space limitations, we use APP usage time as an example (with similar results observed for Video View Count). The comparison results are presented in Figure \ref{fig:compare_app}. It can be observed that using single-sided independent modeling causes a shift in the non-treatment output results (i.e., $y_0$) across different treatments. Ideally, the non-treatment output distributions should be the same, this inconsistency introduces noise between treatments, affecting overall performance. Our proposed method uses KL divergence and joint modeling for correction, ensuring that $y_0$ is consistent. This allows for unbiased estimation of each individual's ITE across different treatments.

\subsection{Technical Soundness}
For causal assumptions, our offline evaluation framework was carefully designed to approximate standard assumptions such as unconfoundedness and SUTVA.
Specifically, we constructed a retrospective dataset simulating a randomized controlled trial. Users were divided into pseudo-treatment and control groups, and a re-shuffling strategy was applied to balance key covariates (e.g., historical watch time, content exposure) over a 7-day window. This mitigates selection bias and supports the unconfoundedness assumption by ensuring treatment assignment is conditionally independent of potential outcomes, given observed covariates. Regarding SUTVA, our treatment (i.e., ranked list exposure) is assigned at the user-session level, minimizing interference between units. To reduce position bias and feedback loops, we avoided engagement signals influenced by prior model exposure. Instead, we used logs collected under a stable serving policy to ensure unbiased offline evaluation.

Our model introduces insignificant computational overhead with +0.026\% latency and +0.021\% CPU usage on the online platform, which are not statistically significant. So our model is production-ready with minimal overhead.
Moreover, the tree-structured design is naturally extensible to more treatments and response types. We plan to explore this in future work by scaling to multi-treatment, multi-response settings.

%% file: main.bib
@String{Springer = "Springer-Verlag" }

@inproceedings{bnn,
  title={Learning representations for counterfactual inference},
  author={Johansson, Fredrik and Shalit, Uri and Sontag, David},
  booktitle={International conference on machine learning},
  pages={3020--3029},
  year={2016},
  organization={PMLR}
}

@article{athey2016recursive,
  title={Recursive partitioning for heterogeneous causal effects},
  author={Athey, Susan and Imbens, Guido},
  journal={Proceedings of the National Academy of Sciences},
  volume={113},
  number={27},
  pages={7353--7360},
  year={2016},
  publisher={National Acad Sciences}
}

@inproceedings{snet,
  title={Nonparametric estimation of heterogeneous treatment effects: From theory to learning algorithms},
  author={Curth, Alicia and Van der Schaar, Mihaela},
  booktitle={International Conference on Artificial Intelligence and Statistics},
  pages={1810--1818},
  year={2021},
  organization={PMLR}
}

@inproceedings{PLE,
  title={Progressive layered extraction (ple): A novel multi-task learning (mtl) model for personalized recommendations},
  author={Tang, Hongyan and Liu, Junning and Zhao, Ming and Gong, Xudong},
  booktitle={RecSys},
  pages={269--278},
  year={2020}
}

@article{metalearners,
  title={Metalearners for estimating heterogeneous treatment effects using machine learning},
  author={K{\"u}nzel, S{\"o}ren R and Sekhon, Jasjeet S and Bickel, Peter J and Yu, Bin},
  journal={Proceedings of the national academy of sciences},
  volume={116},
  number={10},
  pages={4156--4165},
  year={2019},
  publisher={National Acad Sciences}
}

@inproceedings{mmoe,
  title={Modeling task relationships in multi-task learning with multi-gate mixture-of-experts},
  author={Ma, Jiaqi and Zhao, Zhe and Yi, Xinyang and Chen, Jilin and Hong, Lichan and Chi, Ed H},
  booktitle={SIGKDD},
  pages={1930--1939},
  year={2018}
}

@article{liu2024benchmarking,
  title={Benchmarking for Deep Uplift Modeling in Online Marketing},
  author={Liu, Dugang and Tang, Xing and Qiao, Yang and Liu, Miao and Sun, Zexu and He, Xiuqiang and Ming, Zhong},
  journal={arXiv preprint arXiv:2406.00335},
  year={2024}
}

@inproceedings{shalit2017estimating,
  title={Estimating individual treatment effect: generalization bounds and algorithms},
  author={Shalit, Uri and Johansson, Fredrik D and Sontag, David},
  booktitle={International conference on machine learning},
  pages={3076--3085},
  year={2017},
  organization={PMLR}
}

@article{radcliffe2011real,
  title={Real-world uplift modelling with significance-based uplift trees},
  author={Radcliffe, Nicholas J and Surry, Patrick D},
  journal={White Paper TR-2011-1, Stochastic Solutions},
  pages={1--33},
  year={2011}
}

@inproceedings{ganite,
  title={GANITE: Estimation of individualized treatment effects using generative adversarial nets},
  author={Yoon, Jinsung and Jordon, James and Van Der Schaar, Mihaela},
  booktitle={International conference on learning representations},
  year={2018}
}

@inproceedings{efin,
  title={Explicit feature interaction-aware uplift network for online marketing},
  author={Liu, Dugang and Tang, Xing and Gao, Han and Lyu, Fuyuan and He, Xiuqiang},
  booktitle={Proceedings of the 29th ACM SIGKDD Conference on Knowledge Discovery and Data Mining},
  pages={4507--4515},
  year={2023}
}

@article{dragonnet,
  title={Adapting neural networks for the estimation of treatment effects},
  author={Shi, Claudia and Blei, David and Veitch, Victor},
  journal={Advances in neural information processing systems},
  volume={32},
  year={2019}
}

@article{criteo,
  title={A large scale benchmark for individual treatment effect prediction and uplift modeling},
  author={Diemert, Eustache and Betlei, Artem and Renaudin, Christophe and Amini, Massih-Reza and Gregoir, Th{\'e}ophane and Rahier, Thibaud},
  journal={arXiv preprint arXiv:2111.10106},
  year={2021}
}

@inproceedings{descn,
  title={DESCN: Deep entire space cross networks for individual treatment effect estimation},
  author={Zhong, Kailiang and Xiao, Fengtong and Ren, Yan and Liang, Yaorong and Yao, Wenqing and Yang, Xiaofeng and Cen, Ling},
  booktitle={Proceedings of the 28th ACM SIGKDD Conference on Knowledge Discovery and Data Mining},
  pages={4612--4620},
  year={2022}
}

@inproceedings{cigf,
  title={Compressed Interaction Graph based Framework for Multi-behavior Recommendation},
  author={Guo, Wei and Meng, Chang and Yuan, Enming and He, Zhicheng and Guo, Huifeng and Zhang, Yingxue and Chen, Bo and Hu, Yaochen and Tang, Ruiming and Li, Xiu and others},
  booktitle={Proceedings of the ACM Web Conference 2023},
  pages={960--970},
  year={2023}
}

@inproceedings{TensorFlow,
  title={$\{$TensorFlow$\}$: a system for $\{$Large-Scale$\}$ machine learning},
  author={Abadi, Mart{\'\i}n and Barham, Paul and Chen, Jianmin and Chen, Zhifeng and Davis, Andy and Dean, Jeffrey and Devin, Matthieu and Ghemawat, Sanjay and Irving, Geoffrey and Isard, Michael and others},
  booktitle={12th USENIX symposium on operating systems design and implementation (OSDI 16)},
  pages={265--283},
  year={2016}
}

@article{adam,
  title={Adam: A method for stochastic optimization},
  author={Kingma, Diederik P and Ba, Jimmy},
  journal={arXiv preprint arXiv:1412.6980},
  year={2014}
}

@article{flextenet,
  title={On inductive biases for heterogeneous treatment effect estimation},
  author={Curth, Alicia and Van der Schaar, Mihaela},
  journal={Advances in Neural Information Processing Systems},
  volume={34},
  pages={15883--15894},
  year={2021}
}

@inproceedings{gong2020edgerec,
  title={EdgeRec: recommender system on edge in Mobile Taobao},
  author={Gong, Yu and Jiang, Ziwen and Feng, Yufei and Hu, Binbin and Zhao, Kaiqi and Liu, Qingwen and Ou, Wenwu},
  booktitle={Proceedings of the 29th ACM International Conference on Information \& Knowledge Management},
  pages={2477--2484},
  year={2020}
}

@article{yao2021survey,
  title={A survey on causal inference},
  author={Yao, Liuyi and Chu, Zhixuan and Li, Sheng and Li, Yaliang and Gao, Jing and Zhang, Aidong},
  journal={ACM Transactions on Knowledge Discovery from Data (TKDD)},
  volume={15},
  number={5},
  pages={1--46},
  year={2021},
  publisher={ACM New York, NY, USA}
}

@article{cevae,
  title={Causal effect inference with deep latent-variable models},
  author={Louizos, Christos and Shalit, Uri and Mooij, Joris M and Sontag, David and Zemel, Richard and Welling, Max},
  journal={Advances in neural information processing systems},
  volume={30},
  year={2017}
}

@inproceedings{euen,
  title={Addressing exposure bias in uplift modeling for large-scale online advertising},
  author={Ke, Wenwei and Liu, Chuanren and Shi, Xiangfu and Dai, Yiqiao and Philip, S Yu and Zhu, Xiaoqiang},
  booktitle={2021 IEEE International Conference on Data Mining (ICDM)},
  pages={1156--1161},
  year={2021},
  organization={IEEE}
}

@article{neyman_rubin,
  title={Causal inference using potential outcomes: Design, modeling, decisions},
  author={Rubin, Donald B},
  journal={Journal of the American Statistical Association},
  volume={100},
  number={469},
  pages={322--331},
  year={2005},
  publisher={Taylor \& Francis}
}

@inproceedings{gong2022real,
  title={Real-time short video recommendation on mobile devices},
  author={Gong, Xudong and Feng, Qinlin and Zhang, Yuan and Qin, Jiangling and Ding, Weijie and Li, Biao and Jiang, Peng and Gai, Kun},
  booktitle={Proceedings of the 31st ACM international conference on information \& knowledge management},
  pages={3103--3112},
  year={2022}
}

@article{bica2020estimating,
  title={Estimating the effects of continuous-valued interventions using generative adversarial networks},
  author={Bica, Ioana and Jordon, James and van der Schaar, Mihaela},
  journal={Advances in Neural Information Processing Systems},
  volume={33},
  pages={16434--16445},
  year={2020}
}

@inproceedings{xu2022learning,
  title={Learning discriminative representation base on attention for uplift},
  author={Xu, Guoqiang and Yin, Cunxiang and Zhang, Yuchen and Li, Yuncong and He, Yancheng and Cai, Jing and Wei, Zhongyu},
  booktitle={Pacific-Asia Conference on Knowledge Discovery and Data Mining},
  pages={200--211},
  year={2022},
  organization={Springer}
}

@inproceedings{he2024rankability,
  title={Rankability-enhanced Revenue Uplift Modeling Framework for Online Marketing},
  author={He, Bowei and Weng, Yunpeng and Tang, Xing and Cui, Ziqiang and Sun, Zexu and Chen, Liang and He, Xiuqiang and Ma, Chen},
  booktitle={Proceedings of the 30th ACM SIGKDD Conference on Knowledge Discovery and Data Mining},
  pages={5093--5104},
  year={2024}
}

@inproceedings{huang2024entire,
  title={Entire Chain Uplift Modeling with Context-Enhanced Learning for Intelligent Marketing},
  author={Huang, Yinqiu and Wang, Shuli and Gao, Min and Wei, Xue and Li, Changhao and Luo, Chuan and Zhu, Yinhua and Xiao, Xiong and Luo, Yi},
  booktitle={Companion Proceedings of the ACM on Web Conference 2024},
  pages={226--234},
  year={2024}
}

@inproceedings{sun2024m,
  title={M 3 TN: Multi-Gate Mixture-of-Experts Based Multi-Valued Treatment Network for Uplift Modeling},
  author={Sun, Zexu and Chen, Xu},
  booktitle={ICASSP 2024-2024 IEEE International Conference on Acoustics, Speech and Signal Processing (ICASSP)},
  pages={5065--5069},
  year={2024},
  organization={IEEE}
}

@inproceedings{meng2023parallel,
  title={Parallel knowledge enhancement based framework for multi-behavior recommendation},
  author={Meng, Chang and Zhai, Chenhao and Yang, Yu and Zhang, Hengyu and Li, Xiu},
  booktitle={Proceedings of the 32nd ACM International Conference on Information and Knowledge Management},
  pages={1797--1806},
  year={2023}
}

@inproceedings{nandy2023generalized,
  title={Generalized causal tree for uplift modeling},
  author={Nandy, Preetam and Yu, Xiufan and Liu, Wanjun and Tu, Ye and Basu, Kinjal and Chatterjee, Shaunak},
  booktitle={2023 IEEE International Conference on Big Data (BigData)},
  pages={788--798},
  year={2023},
  organization={IEEE}
}

@inproceedings{zhou2023direct,
  title={Direct heterogeneous causal learning for resource allocation problems in marketing},
  author={Zhou, Hao and Li, Shaoming and Jiang, Guibin and Zheng, Jiaqi and Wang, Dong},
  booktitle={Proceedings of the AAAI Conference on Artificial Intelligence},
  volume={37},
  number={4},
  pages={5446--5454},
  year={2023}
}

@article{wei2024multi,
  title={Multi-Treatment Multi-Task Uplift Modeling for Enhancing User Growth},
  author={Wei, Yuxiang and Qiu, Zhaoxin and Li, Yingjie and Sun, Yuke and Li, Xiaoling},
  journal={arXiv preprint arXiv:2408.12803},
  year={2024}
}

@article{liu2009learning,
  title={Learning to rank for information retrieval},
  author={Liu, Tie-Yan and others},
  journal={Foundations and Trends{\textregistered} in Information Retrieval},
  volume={3},
  number={3},
  pages={225--331},
  year={2009},
  publisher={Now Publishers, Inc.}
}

@inproceedings{curth2021nonparametric,
  title={Nonparametric estimation of heterogeneous treatment effects: From theory to learning algorithms},
  author={Curth, Alicia and Van der Schaar, Mihaela},
  booktitle={International Conference on Artificial Intelligence and Statistics},
  pages={1810--1818},
  year={2021},
  organization={PMLR}
}

@article{ricci2021recommender,
  title={Recommender systems: Techniques, applications, and challenges},
  author={Ricci, Francesco and Rokach, Lior and Shapira, Bracha},
  journal={Recommender systems handbook},
  pages={1--35},
  year={2021},
  publisher={Springer}
}

@article{cdum,
  title={Coarse-to-fine Dynamic Uplift Modeling for Real-time Video Recommendation},
  author={Meng, Chang and Zhai, Chenhao and Wang, Xueliang and Liu, Shuchang and Feng, Xiaoqiang and Hu, Lantao and Li, Xiu and Li, Han and Gai, Kun},
  journal={arXiv preprint arXiv:2410.16755},
  year={2024}
}

@inproceedings{wang2015robust,
  title={Robust tree-based causal inference for complex ad effectiveness analysis},
  author={Wang, Pengyuan and Sun, Wei and Yin, Dawei and Yang, Jian and Chang, Yi},
  booktitle={Proceedings of the Eighth ACM International Conference on Web Search and Data Mining},
  pages={67--76},
  year={2015}
}

@article{copf,
  title={Combinatorial Optimization Perspective based Framework for Multi-behavior Recommendation},
  author={Zhai, Chenhao and Meng, Chang and Yang, Yu and Zhang, Kexin and Zhao, Xuhao and Li, Xiu},
  journal={arXiv preprint arXiv:2502.02232},
  year={2025}
}

@inproceedings{li2025contrastive,
  title={Contrastive Prototype Framework for Calibrating Video Recommendation},
  author={Li, Fan and Huang, Jiazhen and Tang, Shisong and Han, Bing and Cao, Huafeng and Sui, Haochen and Xu, Ting and Kang, Xiaoyu},
  booktitle={Proceedings of the 33rd ACM International Conference on Multimedia},
  pages={6373--6382},
  year={2025}
}

@inproceedings{li2024contextual,
  title={Contextual distillation model for diversified recommendation},
  author={Li, Fan and Si, Xu and Tang, Shisong and Wang, Dingmin and Han, Kunyan and Han, Bing and Zhou, Guorui and Song, Yang and Chen, Hechang},
  booktitle={Proceedings of the 30th ACM SIGKDD Conference on Knowledge Discovery and Data Mining},
  pages={5307--5316},
  year={2024}
}

@article{li2025klan,
  title={KLAN: Kuaishou Landing-page Adaptive Navigator},
  author={Li, Fan and Meng, Chang and Fu, Jiaqi and Liu, Shuchang and Zhang, Jiashuo and Zhang, Tianke and Wang, Xueliang and Feng, Xiaoqiang},
  journal={arXiv preprint arXiv:2507.23459},
  year={2025}
}

@inproceedings{tang2025retrieval,
  title={Retrieval Augmented Cross-Domain LifeLong Behavior Modeling for Enhancing Click-through Rate Prediction},
  author={Tang, Xing and Yang, Chaohua and Fu, Yuwen and Ao, Dongyang and Li, Shiwei and Lyu, Fuyuan and Liu, Dugang and He, Xiuqiang},
  booktitle={Proceedings of the 31st ACM SIGKDD Conference on Knowledge Discovery and Data Mining V. 2},
  pages={4891--4900},
  year={2025}
}

@inproceedings{tang2024touch,
  title={Touch the core: Exploring task dependence among hybrid targets for recommendation},
  author={Tang, Xing and Qiao, Yang and Lyu, Fuyuan and Liu, Dugang and He, Xiuqiang},
  booktitle={Proceedings of the 18th ACM Conference on Recommender Systems},
  pages={329--339},
  year={2024}
}

@inproceedings{sun2024end,
  title={End-to-end cost-effective incentive recommendation under budget constraint with uplift modeling},
  author={Sun, Zexu and Yang, Hao and Liu, Dugang and Weng, Yunpeng and Tang, Xing and He, Xiuqiang},
  booktitle={Proceedings of the 18th ACM Conference on Recommender Systems},
  pages={560--569},
  year={2024}
}

@inproceedings{sun2024towards,
  title={Towards Effective and Efficient Multi-valued Treatment Uplift Modeling in Online Marketing},
  author={Sun, Zexu and Liu, Dugang and Tang, Xing and Weng, Yunpeng and He, Xiuqiang},
  booktitle={International Conference on Database Systems for Advanced Applications},
  pages={121--138},
  year={2024},
  organization={Springer}
}

@inproceedings{wu2025aligning,
  title={Aligning and Balancing ID and Multimodal Representations for Recommendation},
  author={Wu, Binrui and Tang, Shisong and Li, Fan and Han, Bing and Meng, Chang and Xiao, Jingyu and Gao, Jiechao},
  booktitle={Proceedings of the 31st ACM SIGKDD Conference on Knowledge Discovery and Data Mining V. 2},
  pages={5029--5038},
  year={2025}
}

@inproceedings{cuicalibrating,
  title={Calibrating Video Watch-time Predictions with Credible Prototype Alignment},
  author={Cui, Chao and Tang, Shisong and Li, Fan and Gao, Jiechao and Chen, Hechang},
  booktitle={Forty-second International Conference on Machine Learning}
}

@inproceedings{weng2023curriculum,
  title={Curriculum modeling the dependence among targets with multi-task learning for financial marketing},
  author={Weng, Yunpeng and Tang, Xing and Chen, Liang and He, Xiuqiang},
  booktitle={proceedings of the 46th International ACM SIGIR Conference on Research and Development in Information Retrieval},
  pages={1914--1918},
  year={2023}
}
